\documentclass{llncs}

\usepackage{graphicx}
\usepackage{algorithm}
\usepackage{algorithmic}
\usepackage{amssymb,amsmath}
\usepackage{amsfonts}
\usepackage{subfigure}
\usepackage{epstopdf}
\usepackage{multirow}

\begin{document}

\title{Sync-and-Burst: Force-Directed Graph Drawing with Uniform Force Magnitudes}

\author{Farshad Ghassemi Toosi\thanks{Corresponding author} and Nikola S. Nikolov}
\institute{Department of CSIS, University of Limerick, Limerick, Ireland \email{\{farshad.toosi,nikola.nikolov\}@ul.ie}}

\date{\today}

\maketitle

\begin{abstract}
We introduce a force-directed algorithm, called Sync-and-Burst, which falls into the category of classical force-directed graph drawing algorithms. A distinct feature in Sync-and-Burst is the use of simplified forces of attraction and repulsion whose magnitude does not depend on the distance between vertices. Instead, magnitudes are uniform throughout the graph at each iteration and monotonically increase as the number of iterations grows. The Sync-and-Burst layouts are always circular in shape with relatively even distribution of vertices throughout the drawing area. We demonstrate that aesthetically pleasing layouts are achieved in $\mathcal{O}(n)$ iterations.
\end{abstract}

\section{Introduction}\label{sec:introduction}

Force-directed graph drawing algorithms have been established as probably the most successful approach towards drawing generic undirected graphs~\cite{Eades2010}. While the classical force-directed algorithms, such as the spring embedder of Eades~\cite{Eades1984}, the algorithm of Fruchterman and Reingold~\cite{FruRei1991}, the barycentric approach~\cite{Tutte1963,DETT1999} and the algorithm of Kamada and Kawai~\cite{KamKaw1989} perform well on relatively small graphs, it has been observed that they do not scale up well. The larger the input graph is, the more likely a force-directed algorithm is to get trapped into a local minimum and produce a poor result~\cite{Kobourov2013}. There has also been recent work on improving force-directed layouts by introducing a third force in addition to the forces of attraction and repulsion between vertices~\cite{BanEppGooTro2013}.

The algorithm introduced in this paper is a force-directed algorithm which follows the general scheme of the spring embedder algorithm. However, we use simplified forces of attraction and repulsion which do not depend on the distance between vertices. Instead, both the magnitude of the attraction force and the magnitude of the repulsion force are uniform throughout the graph at each given iteration, and they monotonically increase (attraction slower than repulsion) as the number of iterations grows. Unlike in other force-directed algorithms, we do not aim at minimising the total energy of the system. Instead, we experimentally demonstrate that aesthetically pleasing layouts are achieved in $\mathcal{O}(n)$ iterations. This approach results in straight-line layouts with a circular shape and generally with a more even distribution of vertices throughout the drawing area when compared to the results of the algorithm of Fruchterman and Reingold. Furthermore, we show some evidence that our algorithm scales up better than the algorithm of Fruchterman and Reingold and is able to find highly symmetrical layouts of some classes of graphs which do not have symmetrical Fruchterman-Reingold layouts. These are graph whose symmetrical layouts would require relatively long edges, e.g. some Hamiltonian graphs including queen graphs, the Wagner graph and the Heawood graph~\cite{Weisstein2015}.

In section~\ref{sec:preliminaries} we introduce the basic terminology necessary for presenting our algorithm. The algorithm is described in detail in section~\ref{sec:algorithm}. Section~\ref{sec:results} presents a comparison between our layouts and Fruchterman-Reingold layouts of the Rome Graphs and a few other selected graphs. Finally, in section~\ref{sec:conclusions} we draw conclusions from this work.

\section{Preliminaries}\label{sec:preliminaries}

In this work we consider undirected graphs. An \emph{undirected graph} $G(V, E)$ consists of a set of $n>0$ vertices $V = \{v_i: 1\le i \le n\}$ and a set of $ m \ge 0$ edges $E=\{e_i: 1 \leq i \leq m\}$, such that each edge is an unordered pair of vertices. Two vertices $v_i, v_j \in V$ are \emph{adjacent} if $\{v_i, v_j\} \in E$. Two edges are \emph{adjacent} if they share an vertex. The adjacency matrix $A=(a)_{ij}$ of $G$ is an $n \times n$ $(0,1)$-matrix such that $a_{ij} = 1$ iff $\{v_i, v_j\} \in E$.

For simplicity, we refer to undirected graphs simply as \emph{graphs} in the remainder of this paper. By a \emph{layout} of a graph we understand a placement, i.e. $x$- and $y$-coordinates, of all vertices within a given 2D drawing area. Assuming forces of attraction and repulsion are defined between pairs of vertices and considering the graph as a mechanical system, a classical force-directed algorithm starts with a random initial layout and then iteratively finds a layout with a minimal total energy, i.e. a layout that corresponds to a mechanical equilibrium of the system. We denote a particular iteration of the algorithm by $t$ with $t \in [1, +\infty]$.

Generally speaking, a force is a vector with a magnitude and direction. We denote the total force of attraction between vertex $v_i$ and its adjacent vertices at iteration $t$ of a force-directed algorithm by $f_a(v_i, t)$. Similarly, $f_r(v_i, t)$ denotes the total force of repulsion between $v_i$ and other vertices at iteration $t$. We use $f_a(v_i,t).x$, $f_a(v_i,t).y$, $f_r(v_i,t).x$, $f_r(v_i,t).y$ to denote the $x$- and $y$-components of $f_a(v_i,t)$ and $f_r(v_i,t)$ in 2D Cartesian space, respectively. Similarly, $v_{i}(t).x$ and $v_{i}(t).y$ represent the $x$- and $y$-coordinates of vertex $v_i$ at iteration $t$, respectively. 

\section{The Sync-and-Burst Algorithm}\label{sec:algorithm}

The algorithm that we propose consists of two phases. The first \emph{sync phase} brings highly interconnected vertices close to each other, while the second \emph{burst phase} spaces vertices out. Thus, we call our algorithm Sync-and-Burst. It is roughly based on the algorithm of Fruchterman and Reingold~\cite{FruRei1991}, but with some very significant differences. In the first two parts of this section we introduce the forces of attraction and repulsion used in Sync-and-Burst and then we give details about the method we use for determining the $x$- and $y$-coordinates at each iteration. The complete algorithm is outlined in section~\ref{sec:pseudocode}.

\subsection{Forces and Total Magnitude}\label{subsec:algorithm:forces}

Similar to Fruchterman-Reingold, we consider forces of attraction between adjacent vertices and forces of repulsion between any pair of vertices. However, at any particular iteration $t$ of Sync-and-Burst, the repulsion between any two vertices has the same uniform magnitude $M(t) > 0$ which is monotonically increasing with the growth of $t$. Similarly, at any particular iteration $t$, the attraction force between any pair of adjacent vertices has the same magnitude which monotonically increases as the number of iterations grows. That is, in comparison to Fruchterman-Reingold we take a simplified approach where the magnitudes of the attraction and the repulsion forces between vertices do not depend on the distance between them.

We want attraction to be the predominant force in the system in the first iterations, making adjacent vertices sync their positions with each other in the layout. Then, as the number of iterations grows, we want the repulsion force to catch up and make vertices go away from each other at an increasing rate. This can be achieved by having the magnitude of the attraction force between a pair of vertices grow slower than the magnitude of the repulsion force. In this work, we set the magnitude of the attraction force to $mM(t)^{0.9}$ (however, the parameter $0.9$ can potentially be tuned to a different value through experimentation).

With this choice of magnitudes, we consider the total magnitude $\mathcal{M}_a(t)$ of attraction and the total magnitude of repulsion $\mathcal{M}_r(t)$ in the system (not the magnitude of the vector sum of forces) as an estimate of how much attraction and how much repulsion there is in the system at a particular iteration $t$. These total magnitudes and their difference $f(t) =  \mathcal{M}_a(t) - \mathcal{M}_r(t)$ are expressed in Equations~\eqref{eq:total_attraction}-\eqref{eq:total_magnitude}. We call $f(t)$ the \emph{total magnitude} function.

\begin{equation}
\label{eq:total_attraction}
\mathcal{M}_a(t) = 2M(t)^{0.9}m^2
\end{equation}

\begin{equation}
\label{eq:total_repulsion}
\mathcal{M}_r(t) = M(t)n(n-1)
\end{equation}

\begin{equation}
\label{eq:total_magnitude}
f(t) = 2M(t)^{0.9}m^2  - M(t)n(n-1)
\end{equation}

\begin{figure}
\centering
\subfigure[$M(t)=t$]{
\includegraphics[width=0.47\textwidth , height =0.2 \textheight]{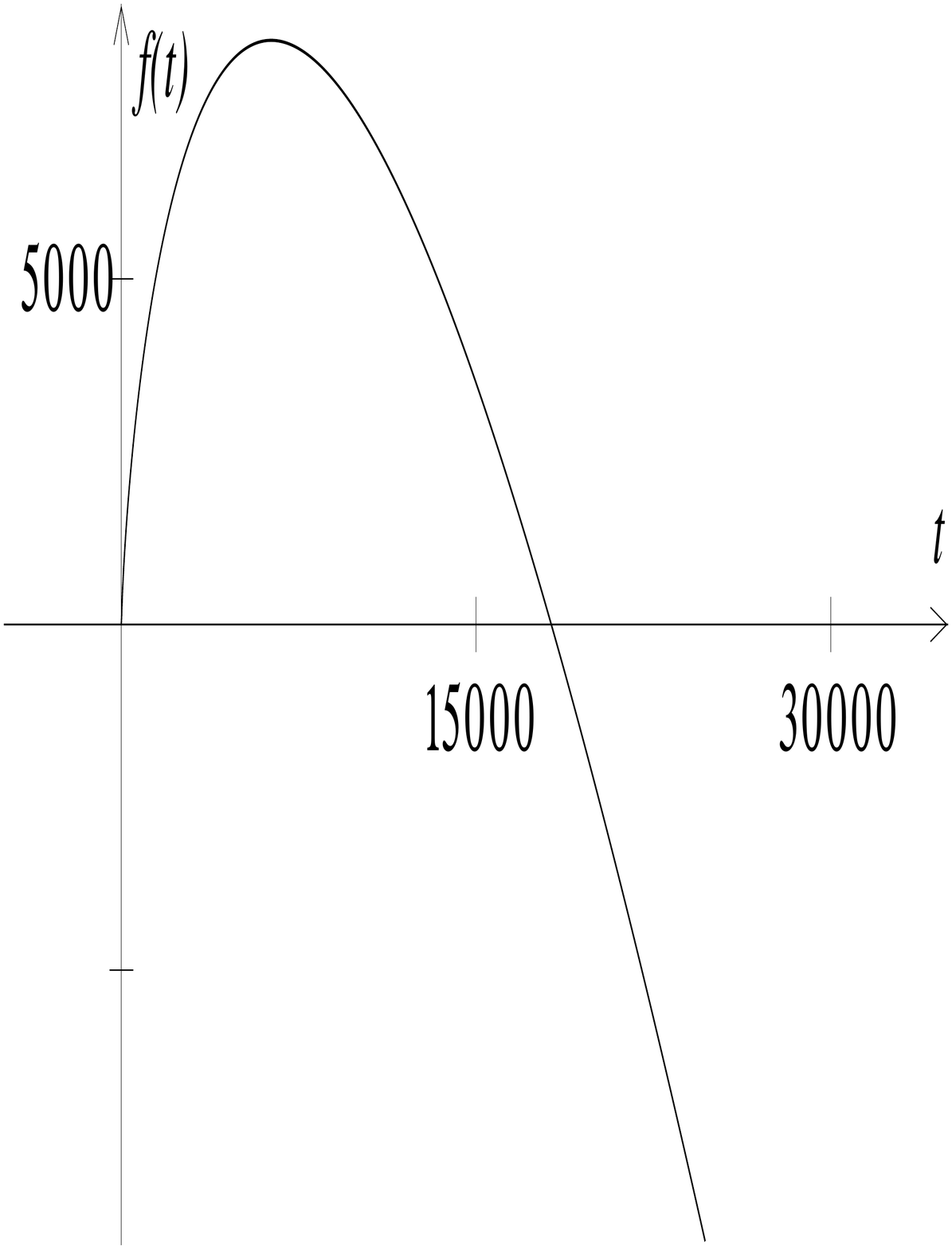}
}
\subfigure[$M(t)=t^{10}$]{
\includegraphics[width=0.47\textwidth, height =0.2 \textheight]{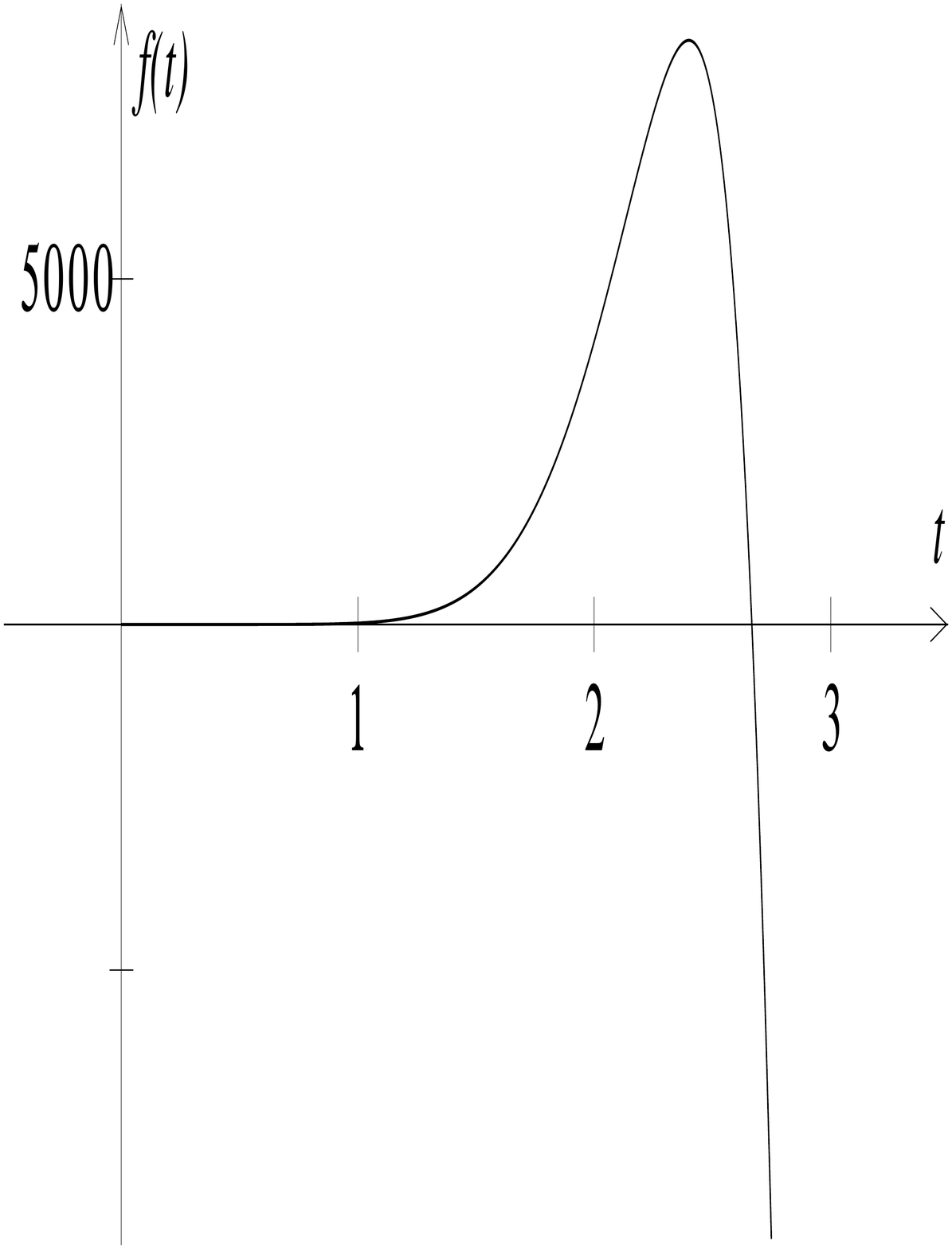}
}
\caption{$f(t)$ (see Equation~\eqref{eq:total_magnitude}) , for a graph with 4 vertices and 4 edges and two different choices of $M(t)$.}
\label{fig:total_magnitude}
\end{figure}

A typical behaviour of $f(t)$ for a monotonically increasing $M(t)$ is plotted in Fig.~\ref{fig:total_magnitude}. The part of the function above the $x$-axis, when attraction is stronger than repulsion, represents the \emph{sync phase} of our algorithm when attraction is the stronger force overall and highly interconnected vertices get relatively close to each other in the layout.  Effectively, in this phase vertices find their approximate placement in the final layout. However, it is likely that some vertices will be placed too close to each other which makes the outcome of this phase aesthetically unacceptable, thus, we continue the iterations. The part of the function below the $x$-axis, when the repulsion becomes stronger overall (and grows increasingly faster than the overall attraction), represents the \emph{burst phase} of our algorithm when vertices are spaced out for achieving an aesthetically pleasing final result.

\subsection{Uniform Force Magnitude} \label{sec:magnitude}

The turning point between the sync and the burst phase is when the total magnitude function $f(t)$ becomes equal to zero, i.e. when the first term $2M(t)^{0.9}m^2$ in Equation~\eqref{eq:total_magnitude} becomes equal to the second term $M(t)n(n-1)$. If this happens for $t = t_p$ then the value of $M(t_p)$ is the one shown in Equation~\eqref{eq:turning_point}. For a graph with $n > 2$ and $m \ge n-1$ (which is a reasonable assumption) $M(t_p) \gg 1$. Thus, by choosing $M(1)<M(t_p)$ and having $M(t)$ increase monotonically, the algorithm starts with syncing the placement of highly interconnected vertices in the first iterations and then, as $M(t)$ becomes larger than $M(t_p)$, vertices are spaced out in the burst phase.

\begin{equation}
\label{eq:turning_point}
M(t_p) = \Bigg(\frac{2m^2}{n(n-1)}\Bigg)^{10}
\end{equation}

Our preliminary experiments with a few graphs, while tuning the parameters of the algorithm, have shown evidence that Sync-and-Burst produces aesthetically pleasing results after a relatively short sync phase followed by a longer burst phase. Thus, if the number of iterations is fixed in advance then we want $M(t)$ to be a function that grows quickly in order to spend fewer iterations in the sync phase than in the burst phase. In order to have our algorithm comparable to other classical force-directed algorithms, we want the total number of iterations to be $\mathcal{O}(n)$. Clearly, by varying the rate at which $M(t)$ grows we can affect how many iterations does the algorithm perform before $M(t)$ reaches the value $M(t_p)$. 

Let the total number of desirable iterations in the sync phase be $sn$ and the total number of desirable iterations in the burst phase be $bn$ where $0 < s < b$. We want $M(t)$ to be a monotonically increasing function that grows fast. Thus, a suitable fit for $M(t)$ would be either a power or an exponential function. The choice we made in this work is to have $M(t) = (t\sqrt[10]{M(t_p)}/sn)^{10}$ which is expressed in terms of $n$ and $m$ in Equation~\eqref{eq:M(t)}. For evaluating Sync-and-Burst we tuned the total number of iterations $(s+b)n$ and the value of the parameter $s$ by experiment as explained in section~\ref{sec:results}.

\begin{equation}
\label{eq:M(t)}
M(t) = \Bigg(\frac{2tm^2}{sn^2(n-1)}\Bigg)^{10}
\end{equation}

\subsection{Vertex Coordinates}\label{subsec:algorithm:coordinates}

Similar to other force-directed algorithms, Sync-and-Burst starts with a random initial layout. At each iteration the coordinates of each vertex $v_i$ are updated based on the total force of attraction $f_a(v_i, t)$ and the total force of repulsion $f_r(v_i, t)$ at $v_i$. Let $\theta_{ij}(t)$ denote the angle between the $x$-axis and the straight line connecting vertices $v_i$ and $v_j$ at iteration $t$. Then we compute the $x$- and $y$-components of $f_a(v_i, t)$ as shown in Equations~\eqref{subeq:fa_x_component} and \eqref{subeq:fa_y_component}, respectively. The value $a_{ij}$ in Equations~\eqref{subeq:fa_x_component} and \eqref{subeq:fa_y_component} is the $ij$-element of the adjacency matrix of the graph. Similarly, the $x$- and $y$-components of $f_r(v_i, t)$ are computed as shown in Equations~\eqref{subeq:fr_x_component} and \eqref{subeq:fr_y_component}. 

\begin{subequations}\label{eq:fa_components}
\begin{align}
f_a(v_i, t).x = mM(t)^{0.9}\sum_{j \neq i}\big(a_{ij}\cos{\theta_{ij}(t-1)}\big)\label{subeq:fa_x_component}\\
f_a(v_i, t).y = mM(t)^{0.9}\sum_{j \neq i}\big(a_{ij}\sin{\theta_{ij}(t-1)}\big)\label{subeq:fa_y_component}
\end{align}
\end{subequations}

\begin{subequations}\label{eq:fr_components}
\begin{align}
f_r(v_i, t).x = -M(t)\sum_{j \neq i}\cos{\theta_{ij}(t-1)}\label{subeq:fr_x_component}\\
f_r(v_i, t).y = -M(t)\sum_{j \neq i}\sin{\theta_{ij}(t-1)}\label{subeq:fr_y_component}
\end{align}
\end{subequations}

\begin{subequations}\label{eq:vertex_coordinates}
\begin{align}
v_{i}(t).x = f_a(v_i, t).x + f_r(v_i, t).x\label{subeq:vertex_x_coordinate}\\
v_{i}(t).y = f_a(v_i, t).y + f_r(v_i, t).y\label{subeq:vertex_y_coordinate}
\end{align}
\end{subequations}

Unlike a typical force-directed algorithm, at each given iteration Sync-and-Burst does not move vertex $v_i$ to a new position starting from its previous position. Instead, the new position of vertex $v_i$ depends on its previous position indirectly. It is influenced by the positions of all vertices relative to $v_i$ at the previous iteration through the angles $\theta_{ij}$ (see Equations~\eqref{subeq:vertex_x_coordinate} and \eqref{subeq:vertex_y_coordinate}). Finally, in order to draw the graph, we normalise the computed coordinates by scaling them to the range $[0, 1]$.

\subsection{The Complete Algorithm} \label{sec:pseudocode}

The complete Sync-and-Burst algorithm is presented as Algorithm~\ref{alg:sync-and-burst}. The body of the outermost \texttt{for}-loop in lines \ref{alg:iteration_start}-\ref{alg:iteration_end} is a single iteration of Sync-and-Burst. In total, we have $(s+b)n$ iterations, where $s$ and $b$ are integer constants which represent the lengths of the sync and burst phases, respectively. The algorithm starts with an initial random layout. Since the first iteration of the algorithm needs the previous value of the uniform force magnitude, i.e. $M(0)$, we set $M(0) = 1/m$ with the intention that we start with a relatively small magnitude below $1$. Then for $t \ge 1$ the value of $M(t)$ is computed according to the formula in Equation~\eqref{eq:M(t)}.

For each vertex $v_i$, the $x$- and $y$-components of the total attraction force applied to it are computed in lines \ref{alg:attraction_start}-\ref{alg:attraction_end}, and the influence of the total repulsion force applied to it is taken into account in lines \ref{alg:repulsion_start}-\ref{alg:repulsion_end}. The values $\theta_{ij}(0)$ in the first iteration of the outermost \texttt{for}-loop are computed from the initial random layout. Finally, we have the coordinates of vertex $v_i$ at iteration $t$ in lines \ref{alg:coordinates_start}-\ref{alg:coordinates_end}. If the layout at iteration $t$ has to be drawn, we scale all coordinates to fit in the interval $[0, 1]$.

\begin{algorithm}
\caption{Sync-and-Burst for graph $G=(V, E)$ with $|V|=n$ and $|E|=m$}
\begin{algorithmic}[1]
\STATE place all vertices at randomly picked positions in the drawing area
\STATE $t \leftarrow 0$
\STATE $M(0) \leftarrow 1/m$
\FOR{$t \leftarrow 1$  to $(s+b)n$}
     \STATE $M(t) \leftarrow (2tm^2/(sn^2(n-1)))^{10}$ \label{alg:iteration_start}
     \FOR{$i \leftarrow 1$  to $n$}
          \STATE $fx_{i} \leftarrow 0$
          \STATE $fy_{i} \leftarrow 0$
          \FOR{$j \leftarrow 1$  to $n$}
               \IF{$\{v_i, v_j\} \in E$} \label{alg:attraction_start}
                    \STATE $fx_{i} \leftarrow fx_{i} + mM(t-1)^{0.9} \cos{\theta_{ij}(t-1)}$
                    \STATE $fy_{i} \leftarrow fy_{i} + mM(t-1)^{0.9} \sin{\theta_{ij}(t-1)}$ 
               \ENDIF \label{alg:attraction_end}
               \IF{$i \neq j$} \label{alg:repulsion_start}
                    \STATE $fx_{i} \leftarrow fx_{i} -  M(t-1) \cos{\theta_{ij}(t-1)}$
                    \STATE $fy_{i} \leftarrow fy_{i} -  M(t-1) \sin{\theta_{ij}(t-1)}$ 
               \ENDIF \label{alg:repulsion_end}
           \ENDFOR
           \STATE $v_i(t).x \leftarrow fx_{i}$ \label{alg:coordinates_start}
           \STATE $v_i(t).y \leftarrow fy_{i}$ \label{alg:coordinates_end} \label{alg:iteration_end}
     \ENDFOR
\ENDFOR
\end{algorithmic}
\label{alg:sync-and-burst}
\end{algorithm}

The computational complexity of Sync-and-Burst is $\mathcal{O}(n^3)$ since the outermost loop is executed $(s+b)n$ times and its body contains two loops, each visiting all vertices. Thus, Sync-and-Burst is comparable to other classical force-directed algorithms, including Fruchterman-Reingold, which have the same computational complexity~\cite{Kobourov2013}. As reported in section~\ref{sec:results}, we have experimentally found that Sync-and-Burst gives aesthetically pleasing results in $20n$ iterations.

\section{Experimental Results and Discussion}\label{sec:results}

We ran both Sync-and-Burst and Fruchterman-Reingold for the Rome Graphs introduced by Di Battista \emph{et al.}~\cite{DiBattista1997}. We used all 11534 Rome Graphs\footnote{available at http://graphdrawing.org/data.html} which have vertex count ranging from $10$ to $100$ vertices. In order to tune the parameters of Sync-and-Burst we ran a pilot study with a few selected graphs. We present the layouts of some of them in section~\ref{sec:selected_graphs}. We discovered that Sync-and-Burst (with the aforementioned choice of the $M(t)$ function) achieves aesthetically pleasing results in $20n$ iterations, where $n$ is the number of vertices, and the length of the sync phase is inversely proportional to the variance of betweenness centralities in the graph. That is, the higher the variance is, the shorter the sync phase needs to be. In particular, we achieved aesthetically pleasing results for all graphs in the pilot study by setting the parameter $s$ to $20$ divided by the standard deviation of betweenness centralities in the graph, but keeping $s$ at most $4$. That is, the sync phase takes at most $20\%$ of all $20n$ iterations and is typically shorter.

We created a Flickr picture gallery\footnote{https://www.flickr.com/photos/133882099@N02} which contains both Sync-and-Burst and Fruchterman-Reingold layouts of selected graphs (including some of the Rome Graphs). 

\subsection{Rome Graphs}\label{sec:rome_graphs}

\begin{figure}
\centering
\subfigure[]{
\includegraphics[width=0.47\textwidth]{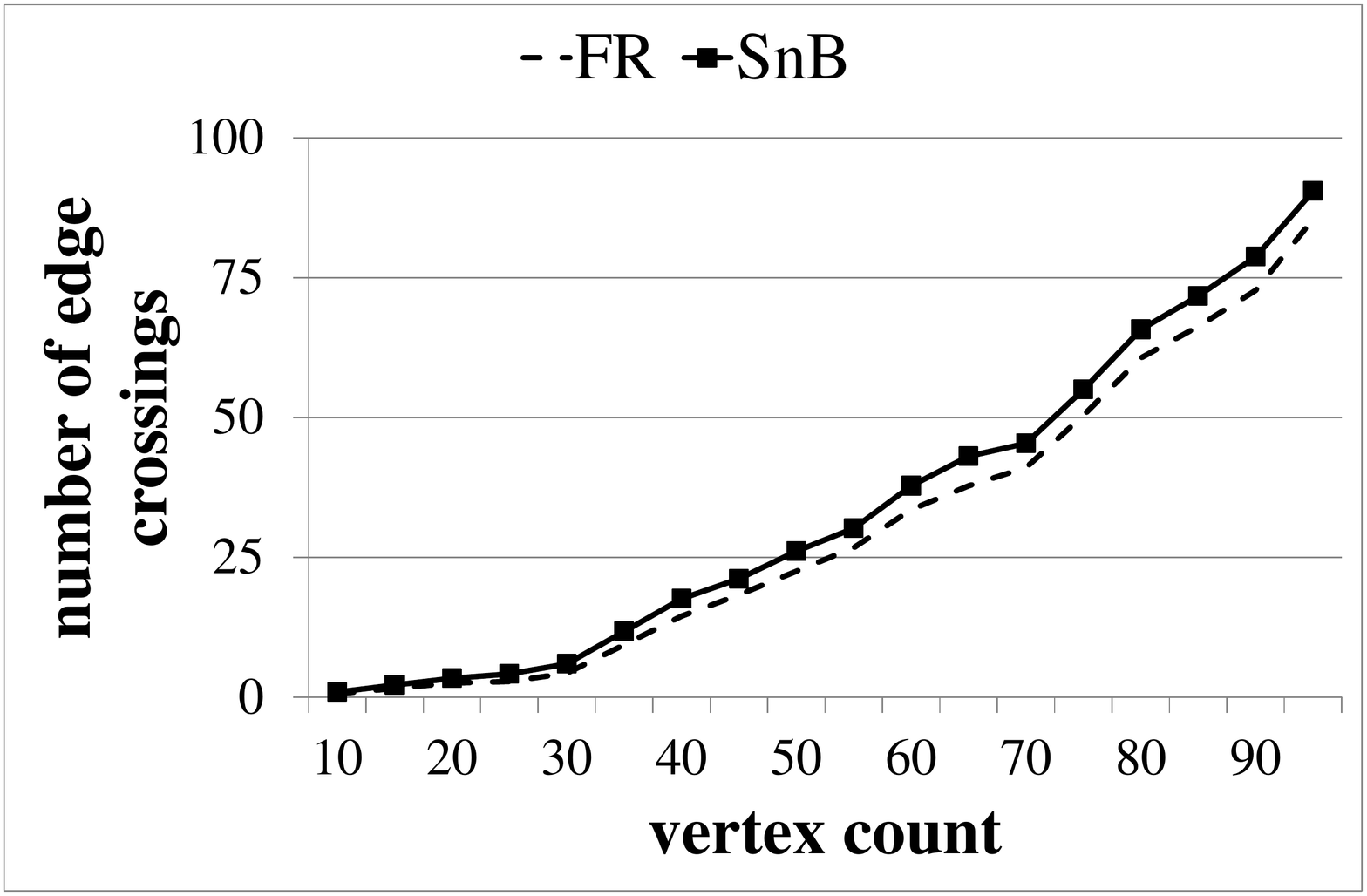}
\label{plot:edge_crossings}
}
\subfigure[]{
\includegraphics[width=0.47\textwidth]{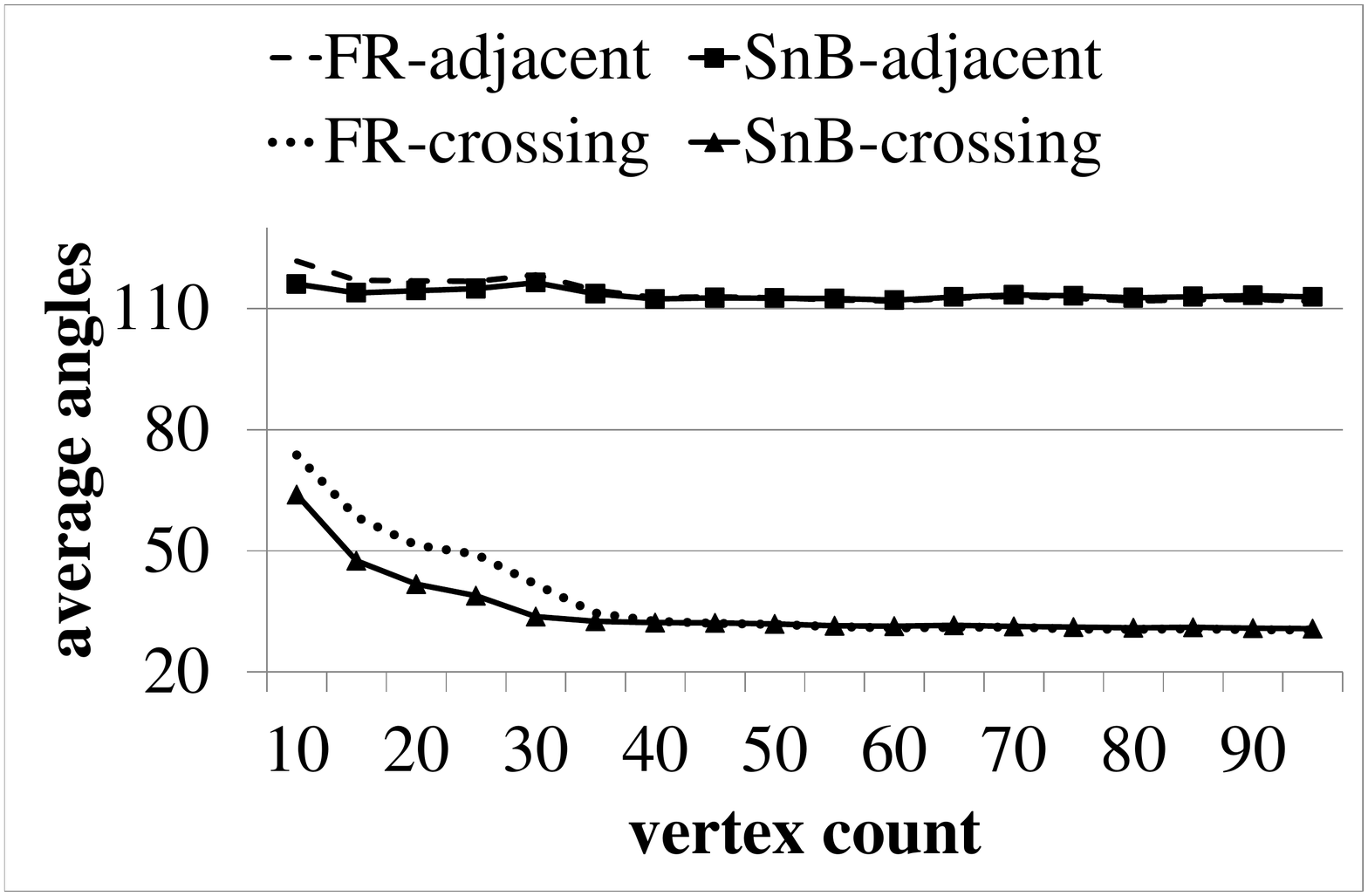}
\label{plot:all_angles}
}
\subfigure[]{
\includegraphics[width=0.47\textwidth]{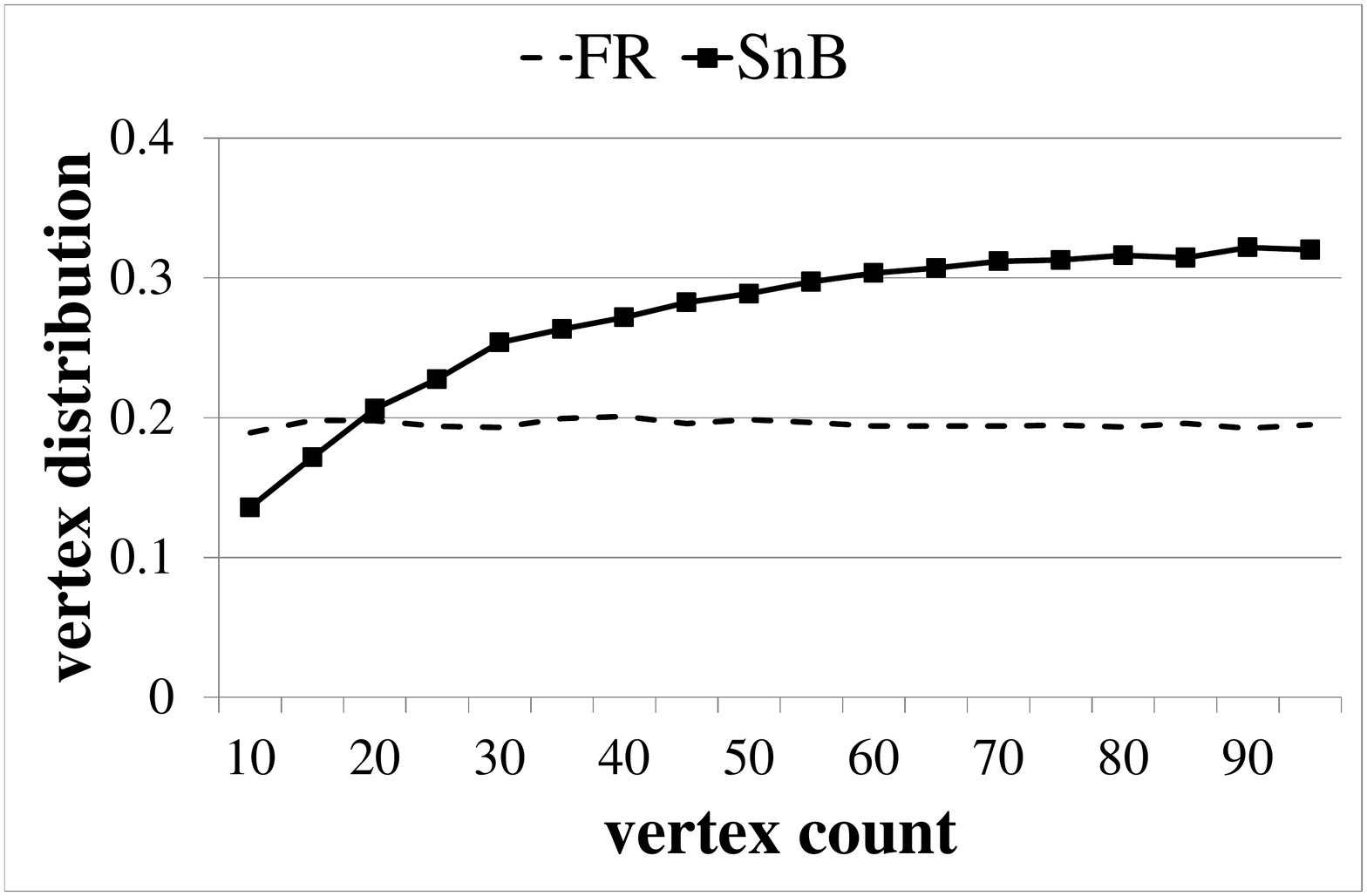}
\label{plot:vertex_distribution}
}
\subfigure[]{
\includegraphics[width=0.47\textwidth]{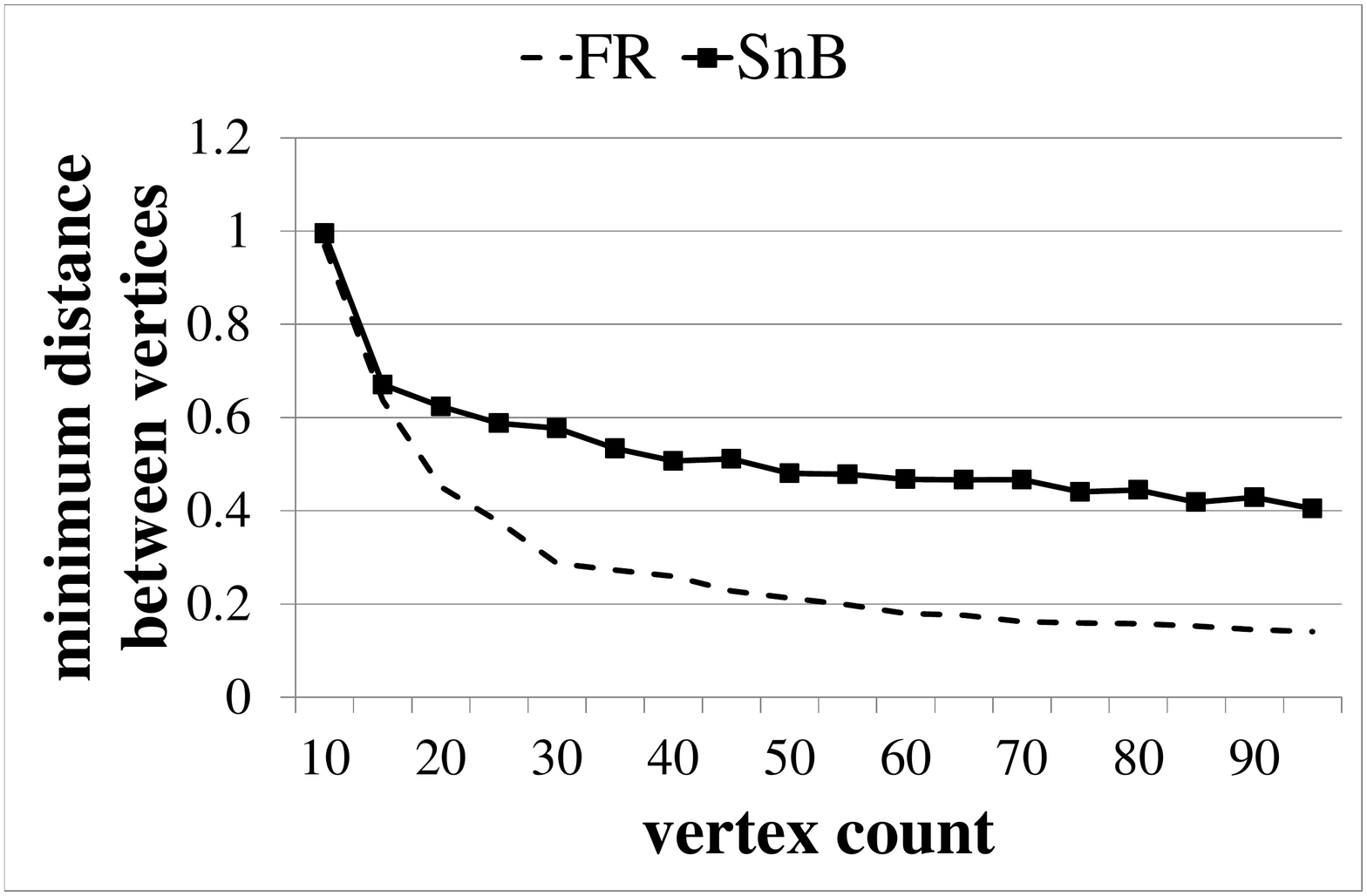}
\label{plot:minimum_distance}
}
\subfigure[]{
\includegraphics[width=0.47\textwidth]{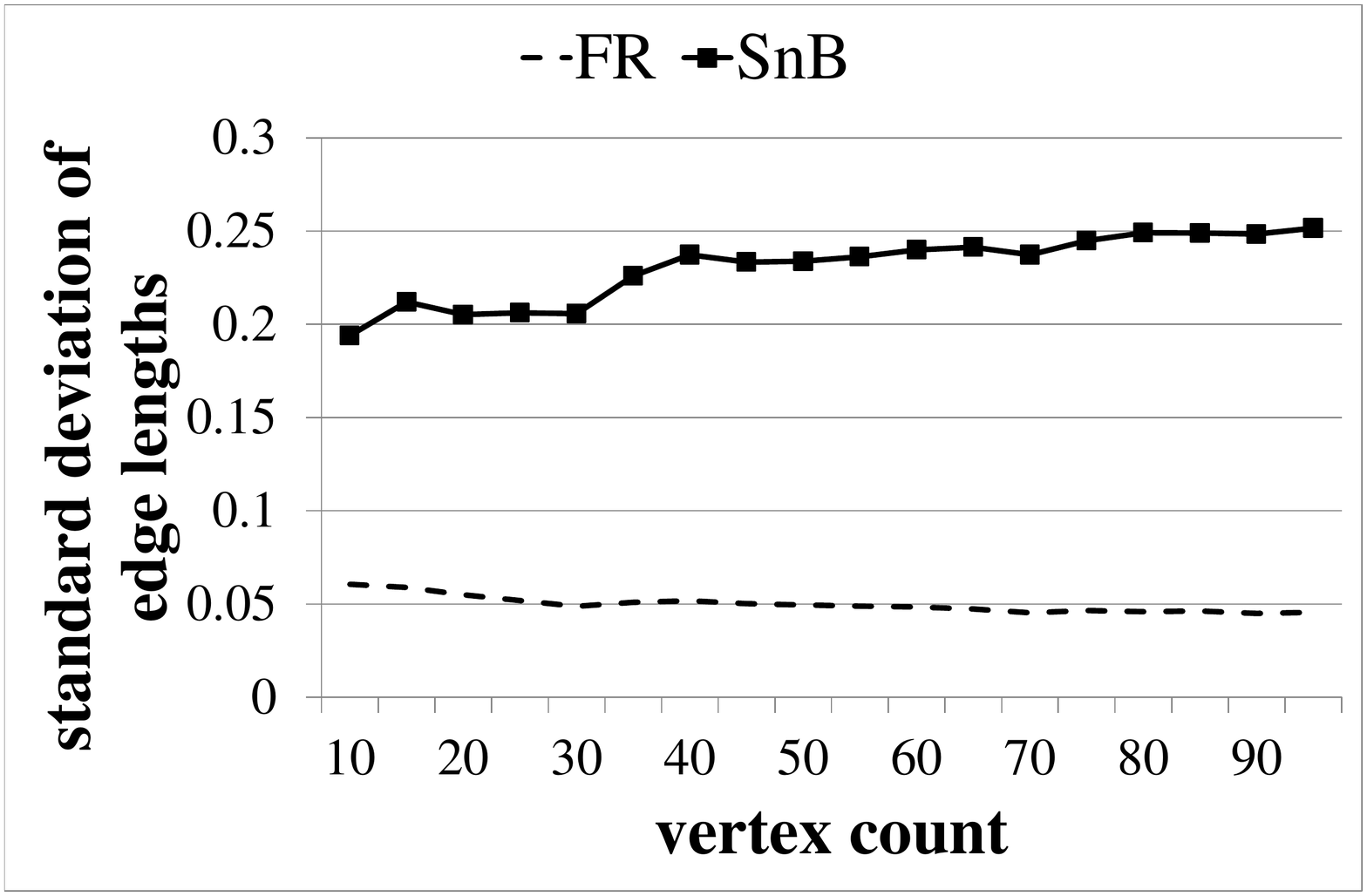}
\label{plot:edge_lengths}
}
\subfigure[]{
\includegraphics[width=0.47\textwidth]{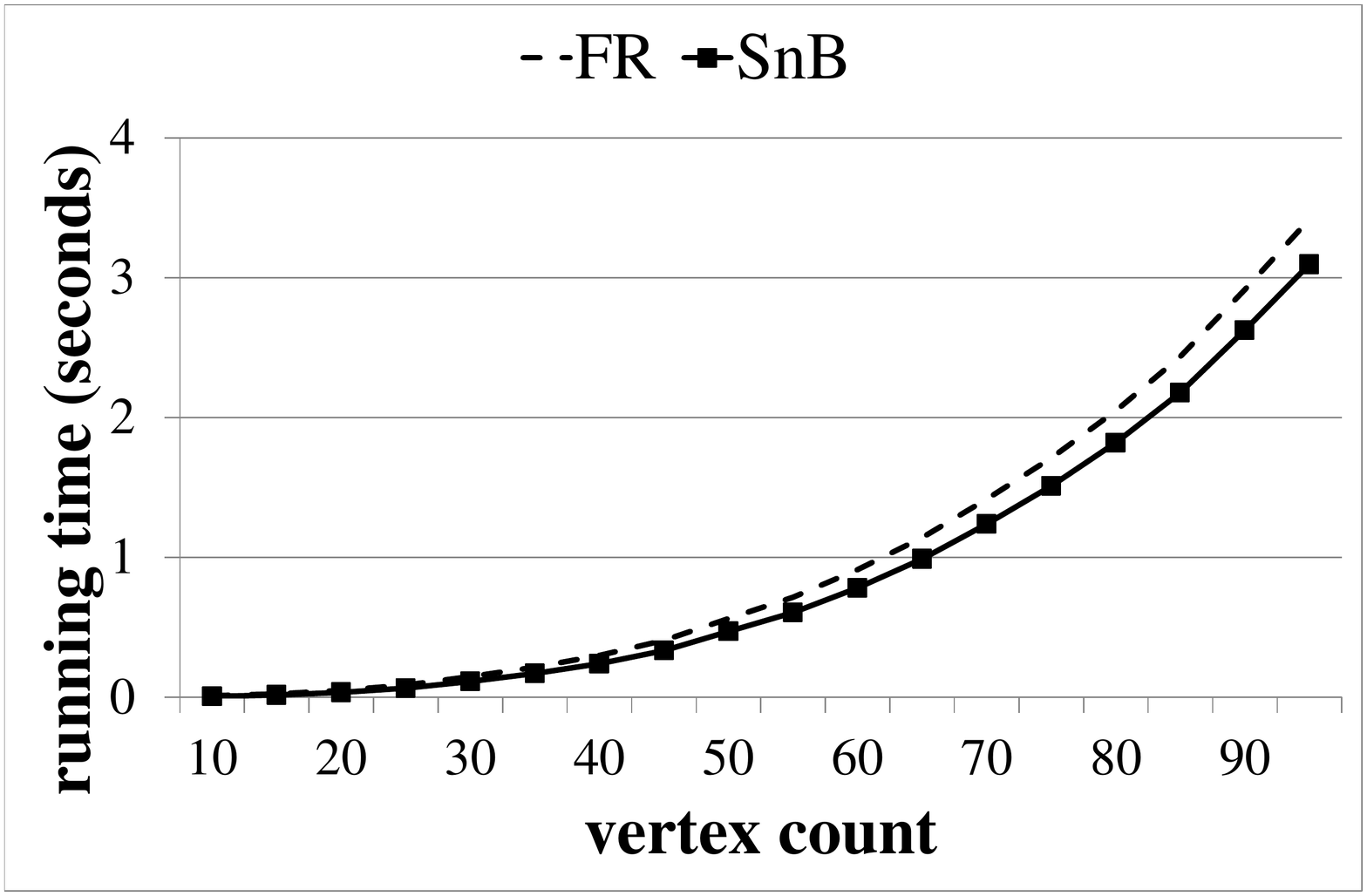}
\label{plot:running_time}
}
\caption{Comparison between Sync-and-Burst and Fruchterman-Reingold layouts of the Rome Graphs: number of edge crossings (a), average angle between any pair of adjacent edges and average crossing angle; the average crossing angle for planar layouts was set to $90^{\circ}$ (b), vertex distribution (c), minimum distance between a pair of vertices multiplied by the number of vertices (d), standard deviation of edge lengths (d), and running time in seconds (d).}
\label{fig:plots}
\end{figure}

For each layout of a Rome graph we recorded a number of characteristics summarised in Fig.~\ref{fig:plots}. These are all standard graph drawing aesthetic criteria such as the number of edge crossings, the average angles between adjacent and crossing edges, the standard deviation of edge lengths and a numerical expression of the distribution of vertices throughout the drawing area. One way of quantifying the distribution of vertices is by finding the minimum distance between two vertices in the layout. The higher this value is the more evenly distributed the vertices are~\cite{LyoMeiRap1998}. We also used a second method of quantifying the distribution of vertices. Let $d^*_i$ be the distance between vertex $v_i$ and the vertex which is closest to it in the layout; let $d^{**}_i$ be the distance from $v_i$ to the closest point on the border of the drawing area; and let $r_i = \min\{d^*_i/2, d^{**}_i\}$. Then, what we call \emph{vertex distribution} $D$ in Fig.~\ref{plot:vertex_distribution} is the ratio of the total area covered by the circles with radii $r_i$ and the actual drawing area $A$ as expressed in Equation~\eqref{eq:distribution}. The closer $D$ is to $1$ the more evenly vertices are distributed throughout the drawing area. In all our experiments, the drawing area is the smallest rectangle that the layout (either Sync-and-Burst or Fruchterman-Reingold) can fit in, scaled (together with the layout) to a rectangle the larger side of which has a unit length.

\begin{equation} 
\label{eq:distribution}
D =\frac{\pi\sum_{i=1}^n{r_i}^2}{A}
\end{equation}

For summarising the results, we partitioned the graph dataset into ``buckets'' according to their vertex count, placing a graph with $n$ vertices into bucket $\lfloor n/5 \rfloor$. The values in Fig.~\ref{fig:plots} are the average for a bucket. The results in Figs.~\ref{plot:edge_crossings} and \ref{plot:all_angles} show that there is no clear winner between Sync-and-Burst and Fruchter\-man-Reingold when it comes to the number of edge crossings and angles between crossing and adjacent edges. While the Fruchterman-Reingold layouts have slightly lower number of edge crossings, the Sync-and-Burst layouts tend to have slightly larger angles between both crossing and adjacent edges for Rome Graphs with vertex count above 55 but the difference is negligible.

The actual difference between the two algorithms can be clearly seen in Figs.~\ref{plot:vertex_distribution}, \ref{plot:minimum_distance} and ~\ref{plot:edge_lengths}. Sync-and-Burst clearly distributes vertices increasingly more evenly throughout the drawing area as the size of the graph grows (see Figs.~\ref{plot:vertex_distribution} and \ref{plot:minimum_distance}). The price that Sync-and-Burst pays for the even distribution of vertices is the larger variance of edge lengths (see Fig.~\ref{plot:edge_lengths}). Classical force-directed algorithms achieve compact drawings with relatively uniform edge length, however, frequently also with large unused (blank) space and entangled layouts of dense subgraphs. Sync-and-Burst, on the other hand, achieves much better utilisation of the drawing area with more even distribution of the vertices by allowing longer edges. The results in Figs.~\ref{plot:vertex_distribution} and \ref{plot:minimum_distance} also suggest that Sync-and-Burst may scale up better than Fruchterman-Reingold as it tends to give increasingly better results in terms of vertex distribution as the size of the graph grows.

Finally, we have compared the running times of Fruchterman-Reingold and Sync-and-Burst on a desktop computer with 8 GB RAM and a 3.20 GHz quad-core CPU. Fig.~\ref{plot:running_time} shows that an iteration of Sync-and-Burst is faster than an iteration of Fruchterman-Reingold. This is something we expected because Sync-and-Burst uses simpler force magnitudes which do not depend on the distance between vertices and thus are faster to compute.

\subsection{Selected Graphs}\label{sec:selected_graphs}

\begin{figure}
\centering
\subfigure[$8 \times 8$ SnB]{
\includegraphics[width=0.22\textwidth]{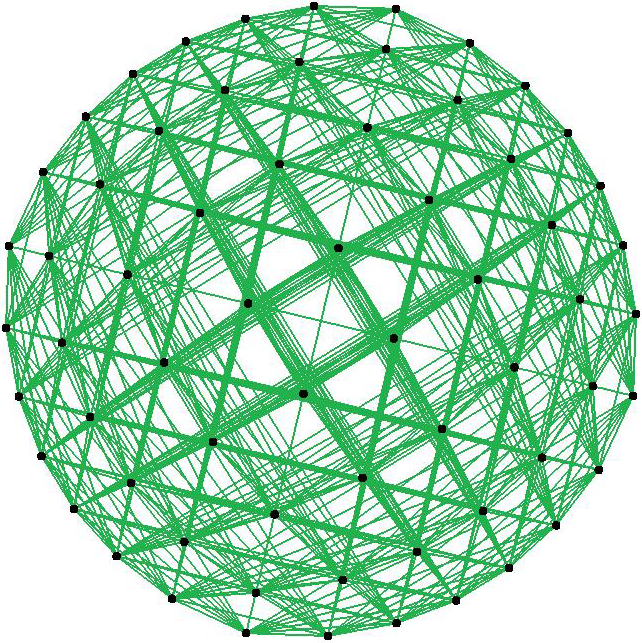}
\label{fig:queen8-8SnB}
}
\subfigure[$8 \times 8$ FR]{
\includegraphics[width=0.22\textwidth]{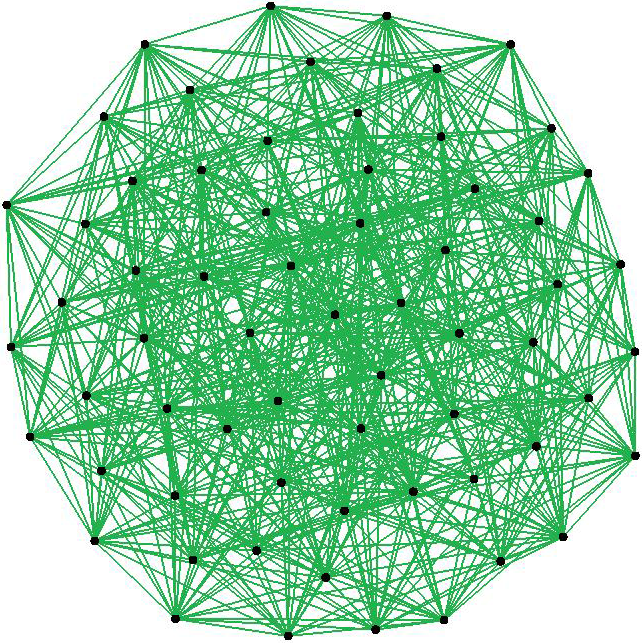}
\label{fig:queen8-8FR}
}
\subfigure[$15 \times 5$ SnB]{
\includegraphics[width=0.22\textwidth]{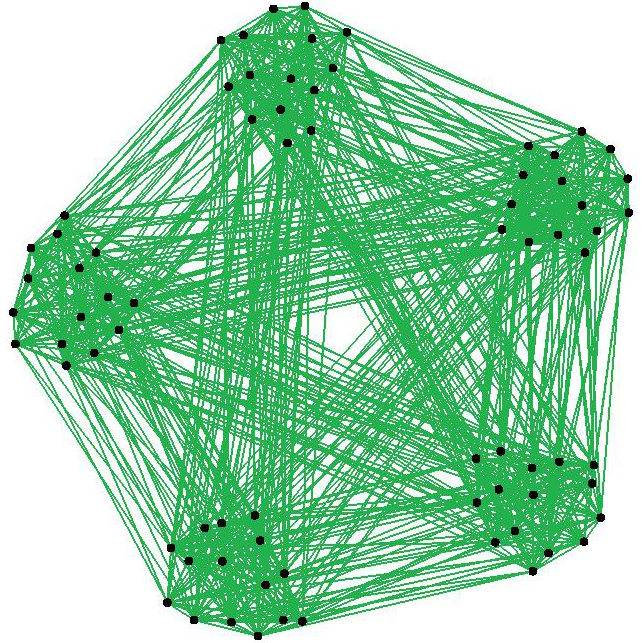}
\label{fig:queen15-5SnB}
}
\subfigure[$15 \times 5$ FR]{
\includegraphics[width=0.22\textwidth]{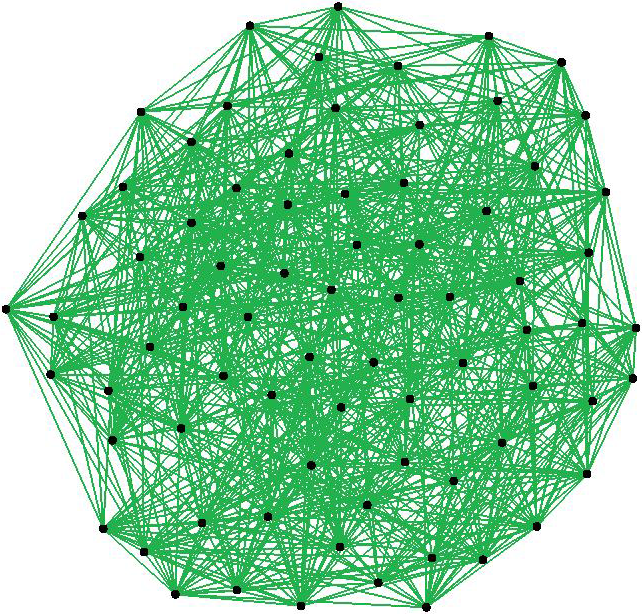}
\label{fig:queen15-5FR}
}
\caption{Sync-and-Burst layouts (a) and (c) compared to Fruchterman-Reingold layouts (b) and (d) of the $8 \times 8$ and the $15 \times 5$ queen graphs. Sync-and-Burst achieves layouts which are clearly more symmetrical.}
\label{fig:example_1}
\end{figure}

\begin{figure}
\centering
\subfigure[graph1 SnB]{
\fbox{\includegraphics[width=0.20\textwidth]{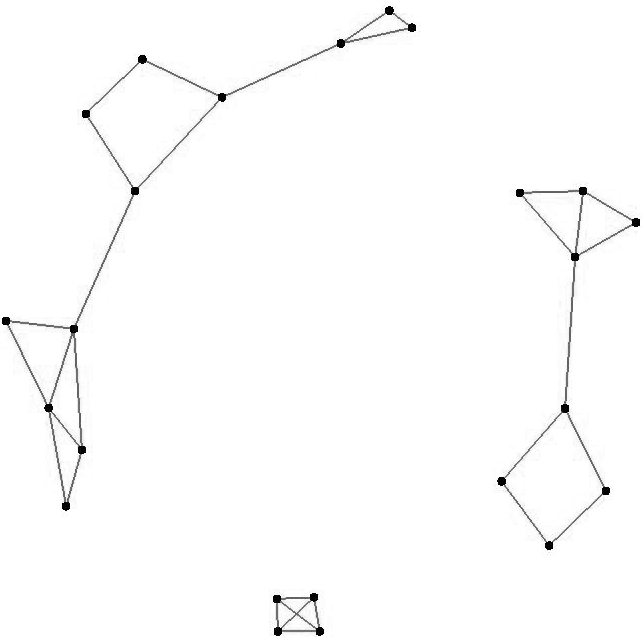}}
\label{fig:disconnectedSnB}
}
\subfigure[graph1 FR]{
\fbox{\includegraphics[width=0.20\textwidth]{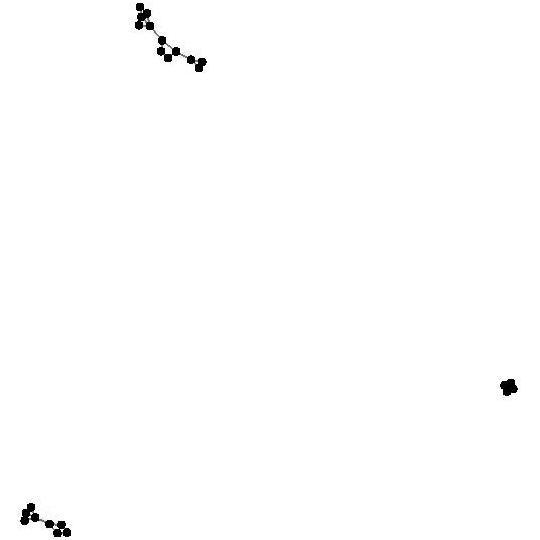}}
\label{fig:disconnectedFR}
}
\subfigure[graph2 SnB]{
\includegraphics[width=0.22\textwidth]{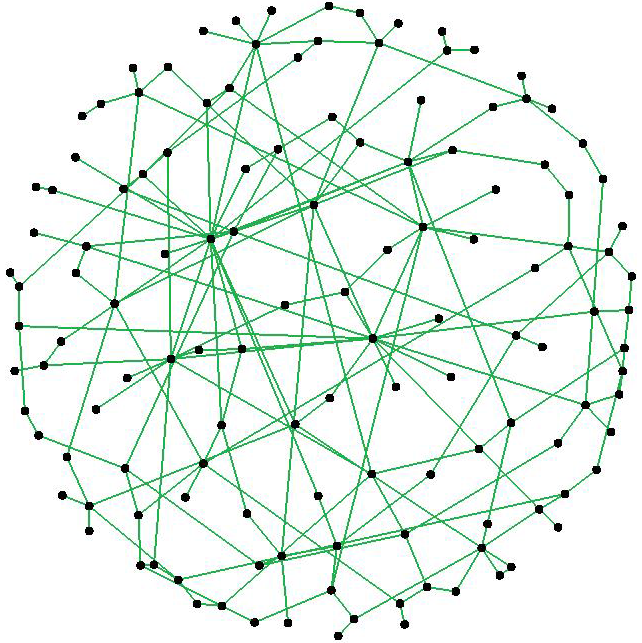}
\label{fig:scalefreeSnB}
}
\subfigure[graph2 FR]{
\includegraphics[width=0.22\textwidth]{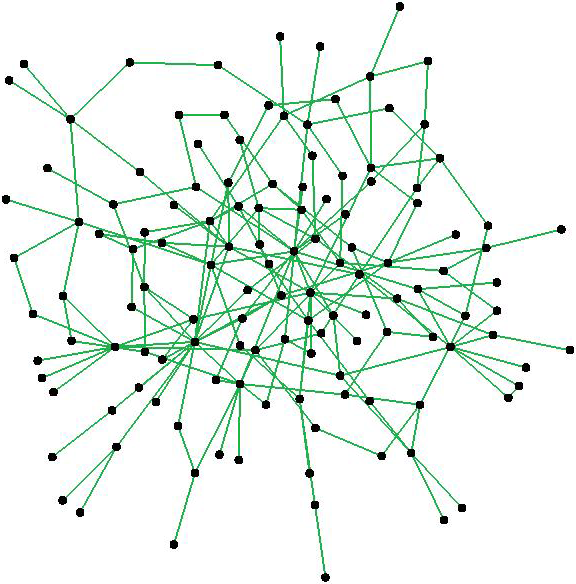}
\label{fig:scalefreeFR}
}\caption{Sync-and-Burst layouts (a) and (c) compared to Fruchterman-Reingold layouts (b) and (d) of a sparse disconnected graph (graph1) and a randomly generated scale-free graph with 130 vertices (graph2).}
\label{fig:example_2}
\end{figure}

\begin{figure}
\centering
\subfigure[Wagner SnB]{
\includegraphics[width=0.22\textwidth]{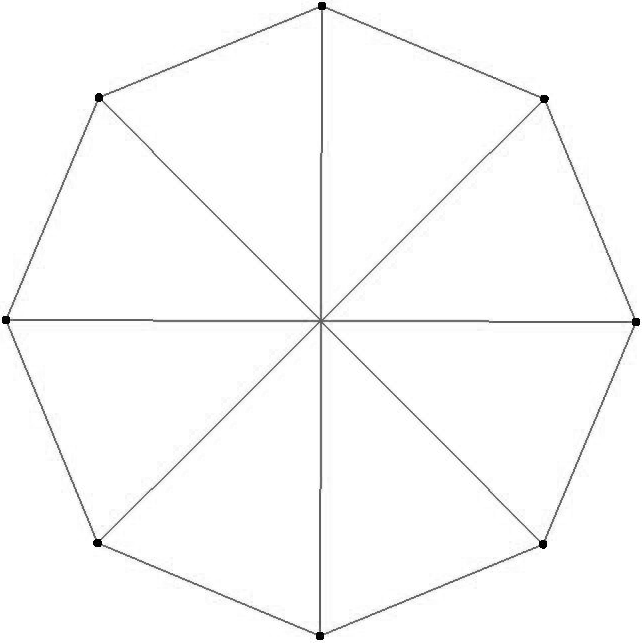}
\label{fig:WagnerSnB}
}
\subfigure[Wagner FR]{
\includegraphics[width=0.22\textwidth]{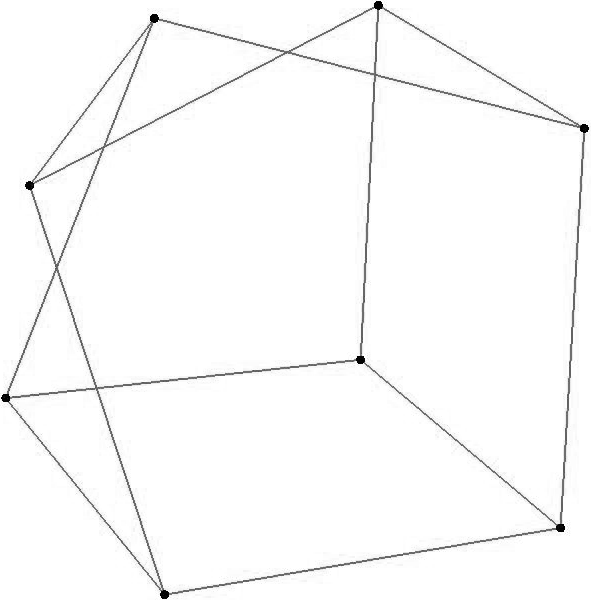}
\label{fig:WagnerFR}
}
\subfigure[Heawood SnB]{
\includegraphics[width=0.22\textwidth]{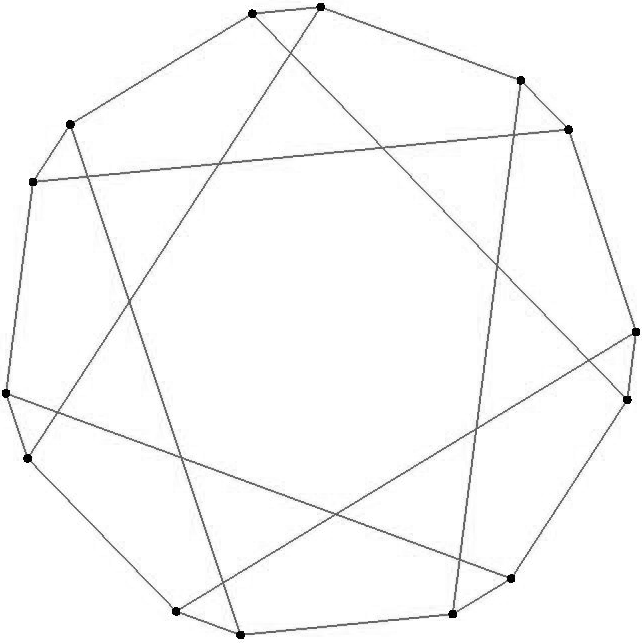}
\label{fig:HeawoodSnB}
}
\subfigure[Heawood FR]{
\includegraphics[width=0.22\textwidth]{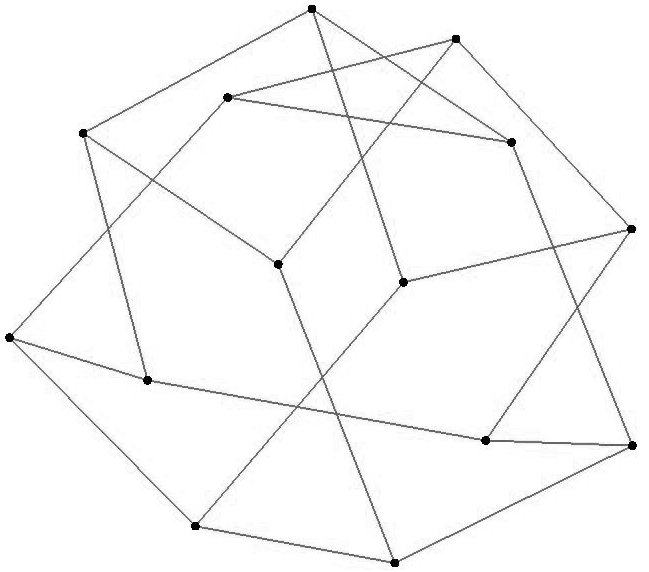}
\label{fig:HeawoodFR}
}\caption{Sync-and-Burst layouts (a) and (c) compared to Fruchterman-Reingold layouts (b) and (d) of a the Wagner graph and the Heawood graph, respectively~\cite{Weisstein2015}.}
\label{fig:example_3}
\end{figure}

In this section we demonstrate the work of Sync-and-Burst on a few selected graphs which we also used in a pilot experimental study to tune parameters of the algorithms. The layouts of $6$ of these graphs are shown Figs.~\ref{fig:example_1}-\ref{fig:example_3}. For comparison, we also present a Fruchterman-Reingold layout of each of these graphs. In the caption of each layout we denote Sync-and-Burst by SnB and Fruchterman-Reingold by FR. The vertex distribution, the number of edge crossings and the standard deviation of edge lengths for the layouts in Figs.~\ref{fig:example_1}-\ref{fig:example_3} are presented in Table~\ref{tbl:selected_graphs}. We counted an edge crossing for each pair of edges which cross, thus in the Sync-and-Burst layout of the Wagner graph (see Fig.~\ref{fig:WagnerSnB}), for example, we counted $6$ edge crossings instead of $1$.

The layouts of the selected graphs confirm the findings in section~\ref{sec:rome_graphs} that Sync-and-Burst distributes vertices evenly throughout the drawing area. Sometimes, this is not in terms of the \emph{vertex distribution} we have measured according to Equation~\eqref{eq:distribution}, but instead in terms of producing a highly symmetrical layout. Such are the symmetrical Sync-and-Burst layouts of the $15 \times 5$ queen graph (see Fig.~\ref{fig:queen15-5SnB}), the Wagner graph (see Fig.~\ref{fig:WagnerSnB}) and the Heawood graph (see Fig.~\ref{fig:HeawoodSnB}) whose Fruchterman-Reingold layouts have higher (i.e. better) vertex distribution according to Equation~\eqref{eq:distribution} (see Table~\ref{tbl:selected_graphs}). Our conjecture is that by producing circular-shape layouts and allowing long edges, Sync-and-Burst is able to highlight symmetries in graphs which do have symmetrical layouts when relatively long edges are permitted. The layouts of the selected graphs also demonstrate that Sync-and-Burst achieves good results for dense and scale-free graphs (see Figs.~\ref{fig:queen8-8SnB}, \ref{fig:queen15-5SnB} and \ref{fig:scalefreeSnB}) which further suggests that it may scale up better than Fruchterman-Reingold. 

\begin{table}
\centering
\caption{Vertex distribution, number of edge crossings and standard deviation of edge lengths for the layouts in Figs.~\ref{fig:example_1}-\ref{fig:example_3}.}
\label{tbl:selected_graphs}
\begin{tabular}{l|cc|cc|cc|cc}
\hline
\multirow{2}{*}{} & \multirow{2}{*}{} & \multirow{2}{*}{} & \multicolumn{2}{|c}{{\bf \begin{tabular}[c]{c}vertex \\ distribution\end{tabular}}} & \multicolumn{2}{|c}{{\bf \begin{tabular}[c]{c}number of \\ edge crossings\end{tabular}}} & \multicolumn{2}{|c}{{\bf \begin{tabular}[c]{c}stdev of\\ edge lengths\end{tabular}}}\\ \cline{4-5} \cline{6-7} \cline{8-9}
{\bf graph}& {\bf $n$}& {\bf $m$} & {\bf SnB} & {\bf FR} & {\bf SnB} & {\bf FR} & {\bf SnB} & {\bf FR} \\ \hline
queen $8 \times 8$~~~      & ~64~ & ~728~ & ~0.480~ & 0.409 & ~31322~ & ~30462~ &  ~0.469~ & ~0.358~ \\ 
queen $15 \times 5$~~~    & ~75~ & 935 & 0.088 & 0.391 & 41950 & 44064 & 0.521 & 0.297 \\ 
graph1                        & ~24~ & 32 & 0.142 & 0.004 & 1 & 1 & 0.095 & 0.009 \\ 
graph2                        & ~130~ & 190 & 0.299 & 0.240 & 287 & 308 & 0.281 & 0.070\\ 
Wagner                       & 8 & 12 & 0.066 & 0.118 & 6 (or 1) & 5 & 0.582 & 0.226 \\ 
Heawood                    & 14 & 21 & 0.038 & 0.217 & 14 & 13 & 0.605 & 0.149 \\ \hline       
\end{tabular}
\end{table}

\section{Conclusions} \label{sec:conclusions}

Sync-and-Burst, the algorithm presented in this paper, falls into the category of classical force-directed algorithms. While following the general schema of the algorithm of Fruchterman and Reingold, Sync-and-Burst uses forces of attraction and repulsion whose magnitude is independent from the distance between vertices. Instead, a uniform magnitude of attraction and a uniform magnitude of repulsion are applied throughout the graph at each iteration and these magnitudes monotonically increase as the number of iterations grows. As a result of this, the Sync-and-Burst layouts are always circular in shape. This could be an advantage for applications where users prefer to see a variety of graphs drawn in the same fashion. Moreover, the experimental study presented in section~\ref{sec:results} demonstrates that Sync-and-Burst distributes vertices more evenly throughout the drawing area than the algorithm of Fruchterman and Reingold, while having similar number of edge crossings and angles between crossing and adjacent edges. Thus, we conclude that Sync-and-Burst distributes vertices more evenly while still achieving a clear and aesthetically pleasing layout. In particular, we achieved very good results for graphs with symmetries in their structure. 

Future work on this algorithm may potentially reveal a better choice for the magnitude function $M(t)$ and the number of iterations, as well as for the ratio between the lengths of the sync and burst phases. In either case, because of the simplicity of the forces, an iteration of Sync-and-Burst can be faster than an iteration of other classical force-directed algorithm (as reported in section~\ref{sec:results}) and we have shown that Sync-and-Burst layouts are aesthetically pleasing in $\mathcal{O}(n)$ iterations. 

\section*{Acknowledgment}

This work is supported by the Irish Research Council (IRC) under project no. GOIPG/2014/938.

\bibliography{ToosiNikolovSnB}
\bibliographystyle{plain}

\end{document}